\documentclass{article}
\usepackage{graphicx}
\usepackage{hyperref}
\usepackage{enumitem}
\usepackage{subcaption}
\usepackage{authblk}

\title{Adapting the serial Alpgen event generator to simulate LHC collisions on millions of parallel threads}

\author{J.T. Childers\footnote{jchilders@anl.gov}}
\author{T.D. Uram}
\author{T.J. LeCompte}
\author{M.E. Papka}
\affil{Argonne National Laboratory, Lemont, IL, USA}
\author{D.P. Benjamin}
\affil{Duke University, Durham, NC, USA}

\begin{document}

\maketitle

\begin{abstract} 
As the LHC moves to higher energies and luminosity, the demand for computing resources increases accordingly and will soon outpace the growth of the Worldwide LHC Computing Grid. To meet this greater demand, event generation Monte Carlo was targeted for adaptation to run on Mira, the supercomputer at the Argonne Leadership Computing Facility. Alpgen is a Monte Carlo event generation application that is used by LHC experiments in the simulation of collisions that take place in the Large Hadron Collider. This paper details the process by which Alpgen was adapted from a single-processor serial-application to a large-scale parallel-application and the performance that was achieved.
\end{abstract}

\section{Introduction}

The US Department of Energy (DOE) continues to invest in scientific computing, driving supercomputers to reach the exaFLOPS scale by the early 2020s. High Energy Physics (HEP) experiments have typically relied on internal resources for computing, with the largest example of this being the Worldwide LHC Computing Grid (WLCG or simply Grid)~\cite{WLCG}. The Grid currently provides $\sim150,000$ concurrent computing cores to each of the experiments ATLAS and CMS, which is equivalent to $1.3$ billion core-hours per year. The average INCITE computing award in 2015 is over $100$ million core-hours per year on the massive parallel supercomputers hosted at current DOE Leadership Computing Facilities. The largest award is $280$ million core-hours. The new machines coming online in 2017 will be a factor of twenty larger. The HEP community is preparing to use these new machines by demonstrating the ability to run on current machines.

This article aims to provide a road map for running existing HEP codes at the highly parallel scales of these machines. Alpgen~\cite{Alpgen}, a FORTRAN-based leading-order multi-parton event-generator for hadronic collisions, is used as an example case. The authors optimize this simulation for running on Mira, the fifth fastest supercomputer in the world, and describe their methods for reaching scales of one million parallel threads in order to facilitate future efforts in HEP.

\paragraph{Motivation}

HEP experiments have traditionally constructed dedicated computing resources needed for simulating and analyzing data produced at accelerator facilities like Fermilab and CERN. However, the computing needs of the LHC experiments at CERN are expected to outpace the resources of the Grid during Run-II. Some of this growth is driven by computationally intensive calculations such as rare processes and matrix elements calculated at next-to-leading order, which are optimal tasks for a supercomputer. Every computing cycle offloaded to supercomputers frees a Grid cycle for other work. In addition, the DOE is pushing to centralize computing resources across HEP experiments and reduce experiment-specific computing resources. Supercomputers offer large computing resources that should become an integral part of the HEP computing strategy.

The Grid is composed of Xeon-class servers and was designed in the era of single-core CPUs with ever-increasing clock speeds. This drove the idea of high throughput computing where performance was measured in the number of serial jobs the system could execute simultaneously. The Grid and software of the experiments using it have been slow to transition to the new multi-core, parallel-processing model of newer processors. Using supercomputers drives applications toward smaller memory footprints and efficient parallel algorithms. The resulting code will easily port to smaller parallel environments on the Grid. In five years, server-class machines will likely transition from 16-core Xeon chips to third-generation Xeon Phi chips with a core count of about one hundred.

\paragraph{Argonne Leadership Computing Facility} 

The Leadership Computing Facility at Argonne National Laboratory (ALCF) hosts Mira, the fifth fastest supercomputer in the world. HEP researchers at Argonne participate in the ATLAS experiment at CERN. This study was performed to assess the usefulness of supercomputers for LHC experiments and the challenges in making code that effectively uses the resources.

Mira is composed of $786,432$ $1600$\,MHz PowerPC A2 cores using the BlueGene/Q architecture, with a peak capability of ten petaFLOPS. Each computing node has 16-cores and 16\,GB of RAM. Each core has four hardware threads. There are no local disks on the computing nodes. Dedicated file-I/O nodes mediate access to the remote GPFS filesystem with one file-I/O node handling file-I/O requests from 128 computing nodes with a peak bandwidth of 1.25\,GB/s. A high-speed 5D Torus network provides 2\,GB/s chip-to-chip communication.

Vesta is a smaller testing and debugging system with $2,048$ nodes, each with 16 PowerPC A2 cores and 16\,GB RAM. In contrast with Mira's compute node-I/O ratio of 128, the I/O infrastructure of Vesta is configured such that 32 computing nodes share a single I/O node.

\paragraph{Alpgen} 

Alpgen\cite{Alpgen} is a FORTRAN-based simulation of multi-parton interactions in hadronic collisions. In this report, Alpgen is used to simulate the production of $W^{\pm}$/$Z$ vector-bosons in association with additional partons (or jets). Alpgen was chosen due to its use in roughly half of the ATLAS experimental results published.

The simulation is composed of three steps: integration, weighted event generation, and unweighting. The integration step calculates the interaction cross-sections of the processes being simulated. The weighted event generation step generates events using the calculated cross-sections and stores these events as their input random seeds and their calculated weight. The unweighting step loops over the weights and regenerates events that pass the unweighting using the stored random seeds. The weighted event generation step is the most computationally intensive of the three, as most weighted events do not pass the unweighting step.

\section{Making Alpgen parallel}

LHC experiments run Alpgen simulations on the Grid by consecutively running the three steps described in the previous section. Each step represents a single execution of the Alpgen code. The integration step uses an inherently serial algorithm accounting for $<1\%$ of the total run-time in the computationally intensive cases and is run on a local computer cluster. The output of the integration can also be used as input to many weighted event generation jobs. The generation of one weighted event is independent of every other event and this is the most computationally intensive of the three steps. These factors made it the first target for parallelization in \emph{Alpgen-v1}. \emph{Alpgen-v2} uses a script to run as a single Mira job the weighted event generation and unweighting steps followed by a file aggregation step. Finally, intermediate files were stored in compute-node memory instead of the filesystem in \emph{Alpgen-v3}.

These developments attempt to minimize the job run-time and improve performance at large scale, while not making changes to the physics algorithms. The ideal case is a code in which run-time is independent of the number of parallel threads; achieving this in light of particular application behaviors and infrastructure is challenging.

\paragraph{Alpgen-v1}

The first parallel version of Alpgen, referred to as \emph{Alpgen-v1}, runs only the weighted event generation step on Mira. The preceding integration step is run on a local computing cluster and the resulting integration output is transferred to Mira. After the weighted event generation job is run on Mira, the per-rank output is aggregated to a single file and returned to the local cluster where the unweighting step is performed.

\emph{Alpgen-v1} uses the Message Passing Interface (MPI)~\cite{MPI} to establish unique random number seeds to ensure that each rank of Alpgen produces different events. When MPI runs parallel instances of Alpgen, it assigns a \emph{rank number}, $n_{rank}$, to each  instance; the rank is a unique number following $0\le n_{rank} < N_{ranks}$ with $N_{ranks}$ being the total number of ranks. Each instance is referred to as a \emph{rank} and MPI provides tools to communicate information between ranks. The use of per-rank random seeds enables \emph{Alpgen-v1} to run at the smallest job size allowed on Mira, 512 nodes.

\begin{figure}
\centering
\begin{subfigure}[b]{0.32\textwidth}
\includegraphics[width=\textwidth,height=0.75\textwidth]{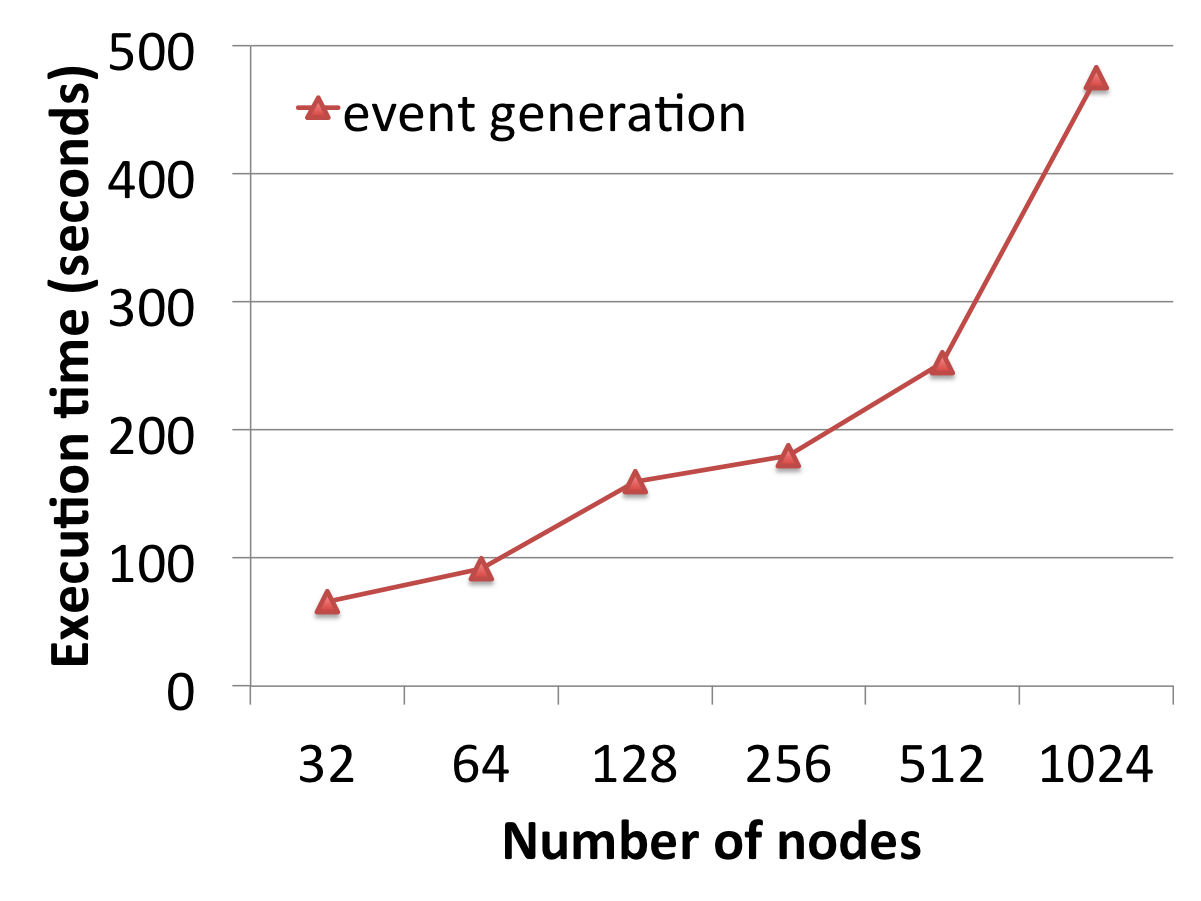}
\caption{\emph{Alpgen-v1}}
\label{fig:alpgenv1_weakscaling}
\end{subfigure}
\begin{subfigure}[b]{0.32\textwidth}
\includegraphics[width=\textwidth,height=0.75\textwidth]{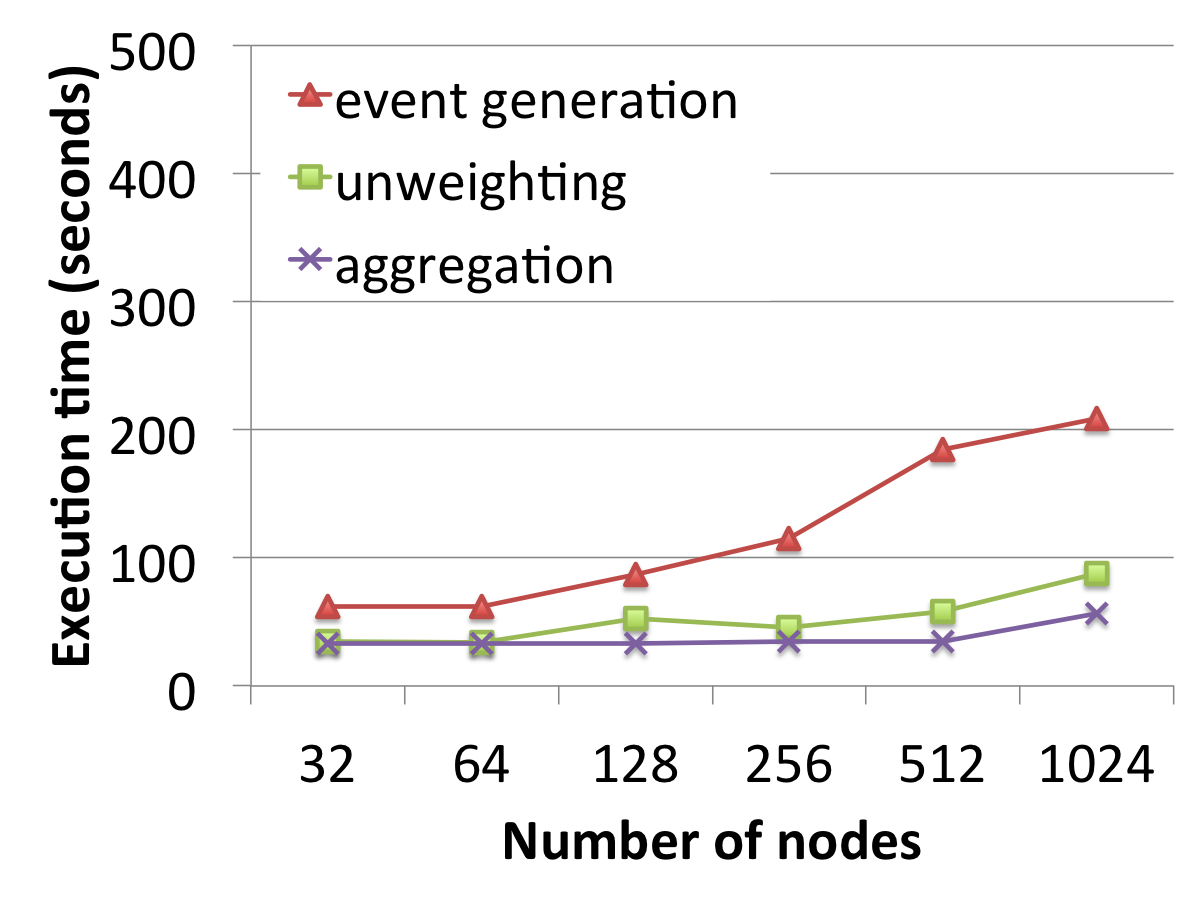}
\caption{\emph{Alpgen-v2}}
\label{fig:alpgenv2_weakscaling}
\end{subfigure}
\begin{subfigure}[b]{0.32\textwidth}
\includegraphics[width=\textwidth,height=0.75\textwidth]{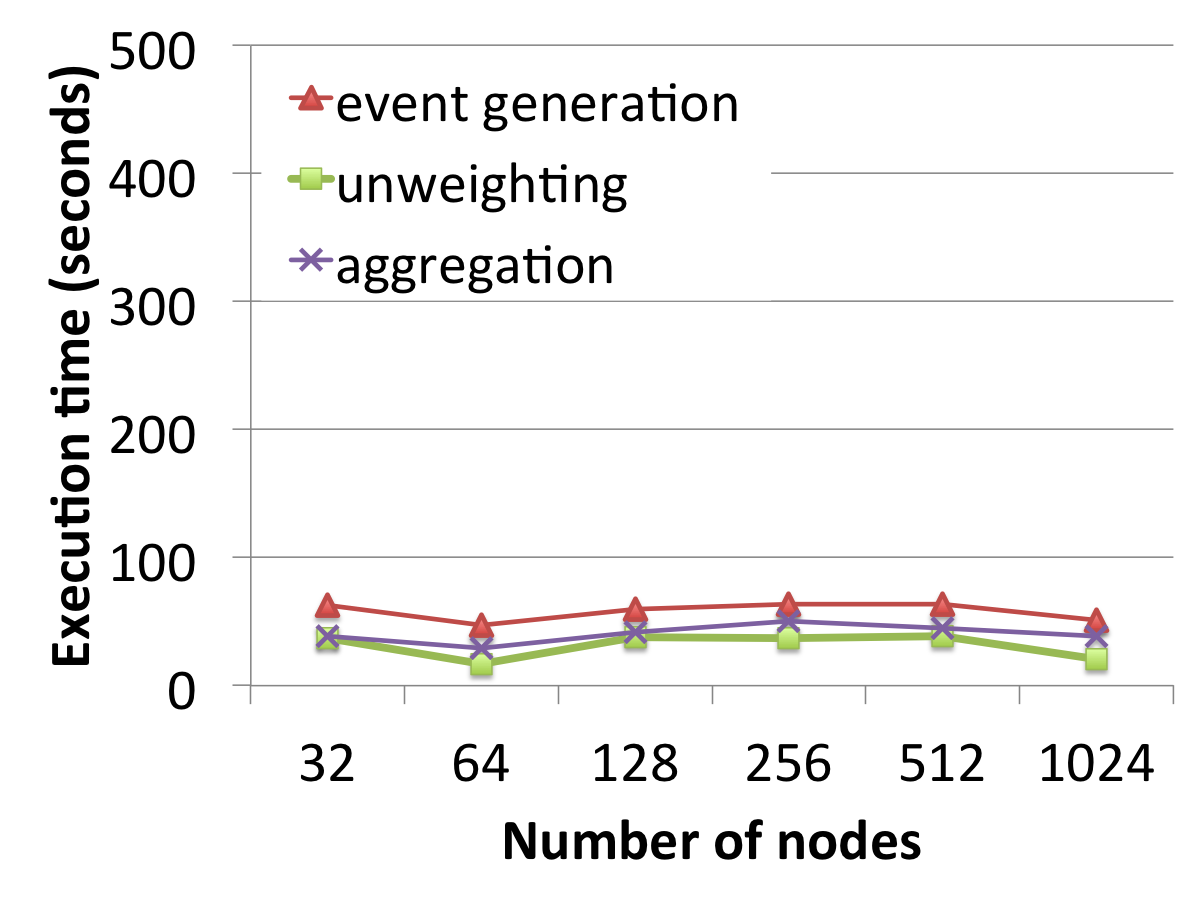}
\caption{\emph{Alpgen-v3}}
\label{fig:alpgenv3_weakscaling}
\end{subfigure}
\caption{Weak scaling of Alpgen versions for the weighted event generation, unweighting, and aggregation phases. These tests were run with 32 ranks per node to simplify the comparison across versions.}
\label{fig:version_comparison}
\end{figure}

Weak scaling runs of \emph{Alpgen-v1} were conducted from 32 to 1,024 nodes on Vesta (a small test system for Mira), with 32 ranks per node. Alpgen was configured to generate 100,000 weighted events per rank. The event generation phase scales poorly with \emph{Alpgen-v1}, as can be seen in Figure \ref{fig:version_comparison}\subref{fig:alpgenv1_weakscaling}. The performance of \emph{Alpgen-v1} is limited by file-per-rank input and output, and per-rank standard output. Also, while it is possible to run as many as 64 threads per node on Mira, due to the large memory footprint of the Alpgen executable, only 32 threads fit in the available memory. 

File-per-rank input and output challenge the filesystem and slow application execution. Every rank of \emph{Alpgen-v1} opens and reads one configuration file, one file containing the Parton Distribution Functions (PDF), and two integration output files, which is about 15\,MB per rank. On one rack of Mira with 1,024 nodes, this is approximately 15\,GB, or roughly 2\,GB per file-I/O node, which is reasonably handled by the aggregate I/O-node bandwidth per rack of 10\,GB/s (1.25\,GB/s per I/O node). This output is, however, written as 128,000 individual files; like other parallel filesystems, Mira's GPFS filesystem scales poorly in the face of file-per-rank inputs and outputs.

Each rank in \emph{Alpgen-v1} writes data to standard output (\texttt{STDOUT}), which is collected by MPI and aggregated at the head node for the job. This flood of small, per-rank output messages stresses the communication infrastructure and prolongs the execution time while messages are collected and flushed to the filesystem. Therefore, \texttt{STDOUT} was limited to a single rank.

In addition to limitations on performance of the weighted event generation on Mira, \emph{Alpgen-v1} is slowed by the preparatory creation of per-rank directories, which is needed to avoid file-I/O collisions between ranks. This step is performed prior to the event generation job running on Mira, but contributes significantly to the overall end-to-end run-time of the weighted event generation step. Relying on the filesystem to isolate threads causes the job run-time to double when the number of threads is doubled, thereby limiting job sizes.

When the job has completed, the weighted events are transferred to the post-processing cluster for unweighting; the transfer and the unweighting tasks are a significant additional cost.

\paragraph{Alpgen-v2}

\emph{Alpgen-v2} overcomes the limitations of \emph{Alpgen-v1} by improving the file access strategy, silencing the per-rank \texttt{STDOUT}, combining the weighted event generation, unweighting, and file aggregation phases into a single job script, and reducing the executable memory footprint allowing 64 Alpgen ranks per Mira node.

Filesystem access is reduced by restricting to one rank the reading of common files (configuration, integration inputs). This leverages the 2GB/s node-to-node network compared to the 1.25\,GB/s file-I/O nodes shared by 128 compute nodes. Unnecessary output files are disabled such as data analysis and run-time status. The directories per rank are no longer necessary, which reduces the per-rank overhead in preprocessing. 

The weighted event generation and unweighting phases are combined using a script to reduce the output data size which greatly reduces the postprocessing time. During weighted event generation, a single thread of \emph{Alpgen-v2} writes two files to the filesystem, a weighted event file and a parameter file. The subsequent unweighting step reads these files from the filesystem as input, and the unweighted events are written to the filesystem. Including the unweighting in the Mira job reduces the output data size: in the case of $W$+5jets, though the unweighted events are larger than weighted events by a factor of 10, the unweighting reduces the number of events by a factor of 10,000 leading to a reduction in data size by a factor of 1000.

Details of the cross-section and total events produced across were stored in per-rank files at the end of the unweighting step in \emph{Alpgen-v1}. \emph{Alpgen-v2} aggregates this information using MPI reduction operations to make a single global output parameter file reducing filesystem access.

An aggregation step is also included in combined script that runs on Mira. This step uses MPI application in which each rank reads the rank-wise unweighted event files and uses MPI collective I/O functions to aggregate the data into a single output file. Reducing the output from one file per rank to a single file reduces stress on the filesystem, allowing the job to finish more quickly, and improving the time for the subsequent transfer to the post-processing cluster.

\texttt{STDOUT} is limited to the $n_{rank}==0$ rank to diminish the overhead of collecting many small messages over MPI, while retaining some output for debugging and logging purposes. Since the output is largely ignored with increasing scale, it can be disabled without impacting application diagnostics.

\emph{Alpgen-v2} reduces the memory footprint to achieve greater node-level parallelism. The original version of Alpgen has a run-time memory occupancy of 203\,MB. Alpgen allocates memory in the data section of the executable for the eight possible PDFs, whereas only one is used during execution. \emph{Alpgen-v2} is altered to include only the single PDF specified by the input configuration, reducing the memory footprint to 10\,MB and allowing for a higher thread count per node. In this configuration, Alpgen is able to run with 64 ranks per node.

Weighted event generation in \emph{Alpgen-v2} exhibits much better scaling, as shown in Figure \ref{fig:version_comparison}\subref{fig:alpgenv2_weakscaling}. It is clear from the figure that the unweighting and aggregation steps perform well up to 512 nodes, with an increase in execution time at 1,024 nodes. The limiting factor in weighted event generation is the writing of one file per rank to the filesystem.

\paragraph{Alpgen-v3}

\emph{Alpgen-v3} improves file access by writing intermediate files to the compute-node persistent memory \cite{BlueGeneQApplicationRedbook} (i.e. RAM-disk) on Blue Gene Q systems. The intermediate files, i.e. weighted event data and parameter files and rank-wise unweighted data files, are stored in the RAM of each compute node. The aggregation step is then used to read the rank-wise unweighted data files from the RAM-disk, and output a single data file to the filesystem. This strategy leverages the many-fold faster compute node memory in place of the filesystem, eliminating round-trip write/read cycles to the filesystem and avoiding the problem of many small file accesses. In the cases of $W$+5jets, $W$+4jets, and $W$+3jets, the average rank produces 188\,KB, 1.2\,MB, and 3.0\,MB of temporary data, respectively, when producing one million weighted events.

Weighted event generation, unweighting, and aggregation all benefit from storing intermediate data files in persistent memory as opposed to the filesystem. Figure \ref{fig:version_comparison}\subref{fig:alpgenv3_weakscaling} shows that the execution time for all three phases is nearly flat up to 1,024 nodes. 

\paragraph{End-to-end workflow run-time analysis}

\begin{figure}
\centering
\includegraphics[width=0.7\textwidth]{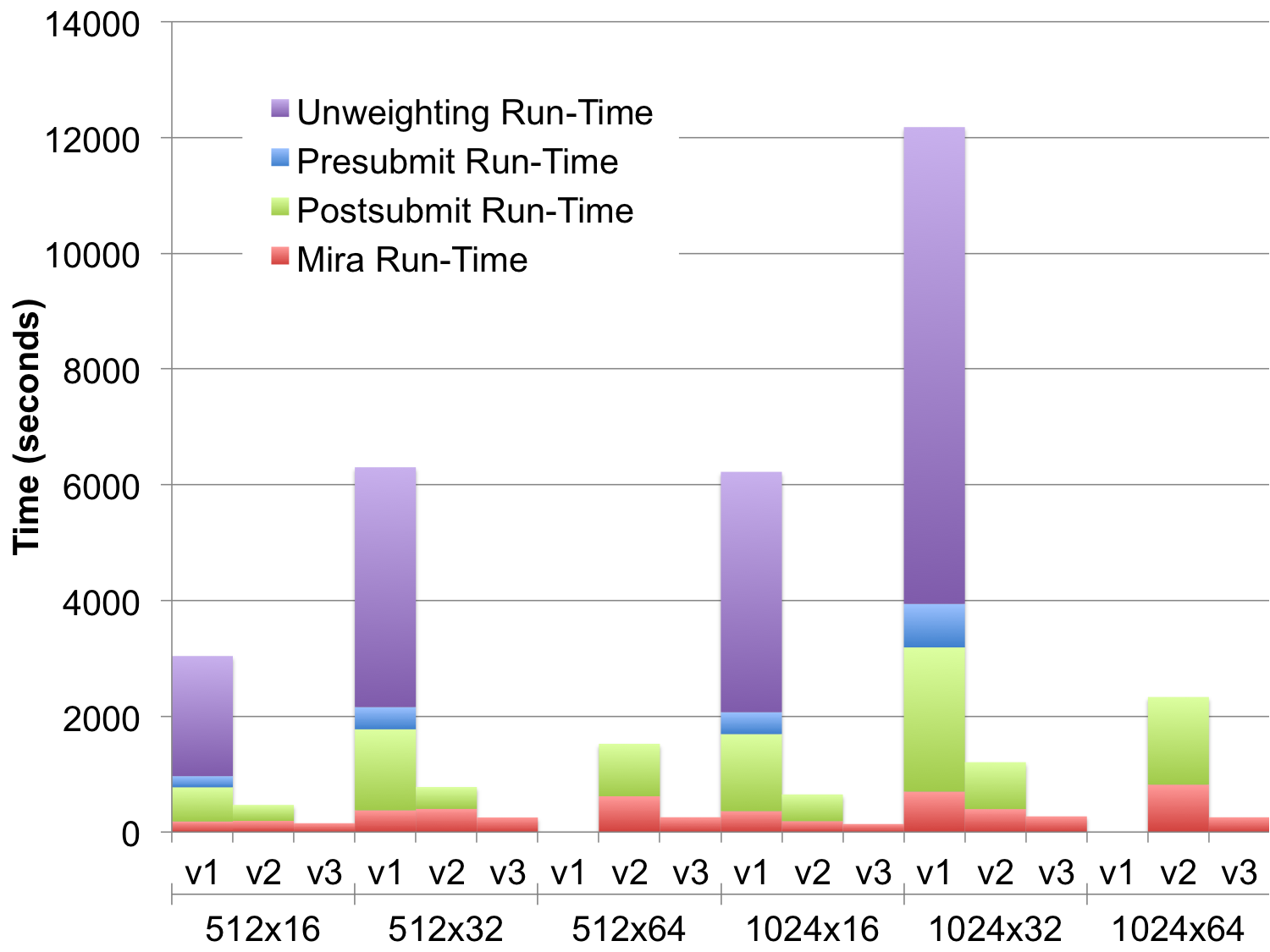}
\caption{Run-times for each Alpgen version for different node and thread per node configurations of Mira jobs. Presubmit run-time includes any file placement or directory creation needed prior to running the Mira job. Postsubmit run-time includes any file aggregation and file or directory cleanup after running the Mira job. The unweighting run-time represents the time taken to run the unweighting task on the weighted events, on the postprocessing cluster.}
\label{fig:run_time_versions}
\end{figure}

The three versions of Alpgen described above were tested with varying numbers of nodes and threads per node to see the improved run-time. The combinations $512\times16$, $512\times32$, $512\times64$, $1024\times16$, $1024\times32$, and $1024\times64$ are shown in Figure~\ref{fig:run_time_versions}, where the different Alpgen versions are represented as \emph{v1}, \emph{v2}, and \emph{v3}. These Alpgen tasks were Z+2jets with each thread producing 350,000 weighted events. The task run-time is divided into these steps:
\begin{itemize}[noitemsep,nolistsep]
\item The Mira run-time represents the actual job run-time on Mira and is the only step running parallel processes.
\item The presubmit run-time represents the time needed to prepare to run on Mira. This only exists for \emph{Alpgen-v1} as it required building the per-rank directory structure described above.
\item The postsubmit run-time represents the time needed to clean up after a run. This only exists for \emph{Alpgen-v1} and \emph{Alpgen-v2}. In the former case, this accounts for aggregating the weighted event generation files and removing the directory structure. In the later case, this accounts for the removal of all the intermediate files which do not need to be kept and are stored in the RAM-disk in \emph{Alpgen-v3}.
\item The unweighting run-time accounts for the time to unweight the weighted events produced in the Mira job for \emph{Alpgen-v1} only. \emph{Alpgen-v2} and \emph{Alpgen-v3} include the unweighting step in the Mira job.
\end{itemize}
The run-time of \emph{Alpgen-v1} grows linearly with job size, as can be seen by comparing runs with a similar number of ranks; for example, 512x16 vs 1024x32. The primary reason for this doubling lies in the time for the presubmit, postsubmit, and unweighting steps to be performed serially on the Mira login nodes or the local cluster. \emph{Alpgen-v1} was not able to run with 64 ranks per node because it exceeded the available system memory at this size.

Overall run-time with \emph{Alpgen-v2} is $>6\times$ less than \emph{Alpgen-v1}, primarily due to the unweighting step being run in parallel after being combined with the weighted event generation step on Mira, in which case it has shrunk to be no longer evident on the chart. The presubmit time has similarly vanished from the graph. Postsubmit time remains and increases correspondingly with the number of ranks.

The run-time of \emph{Alpgen-v3} consists entirely of the time for the weighted event generation, unweighting, and aggregation to run on Mira, finishing approximately $>20\times$ faster than \emph{Alpgen-v1}.

\section{Parallel scaling performance}

The performance of Alpgen on Mira can be characterized by the run-time of identically configured jobs with differing parallel size. Computing time on Mira is charged per core-hour; therefore, the performance metric will be defined as the number of unweighted events, $N_{evt}$, per Mira core-hour, $C$:
\begin{equation}
P = \frac{N_{evt}}{C} = \frac{N_{evt}}{t N_{node} N_{core}}.
\end{equation}
$t$ is the job run-time in hours, $N_{node}$ is the number of nodes in the job, and $N_{core}$ is the number of cores per node which is always $16$ for Mira. $P$ should increase with better performance.

The performance is driven by the hardware architecture. The number of threads per node and the total number of nodes affects performance. Using the total number of threads ($N_{core}\times N_{node}$) is ambiguous because running jobs with 512 nodes at 32 threads per node (written as $512\times32$ henceforth) and $1024\times16$ contain the same total number of threads. However, the performance of these two configurations is not comparable. The thread count per CPU-core is higher in the $512\times32$ job resulting in a higher CPU-load that can increase run-time. If the job is file-I/O intensive, recalling that 128 worker nodes share a single file-I/O node, the $512\times32$ job has a higher throughput on the file-I/O node that can also make job run-times longer.

\begin{figure}
\centering
\includegraphics[width=0.48\textwidth]{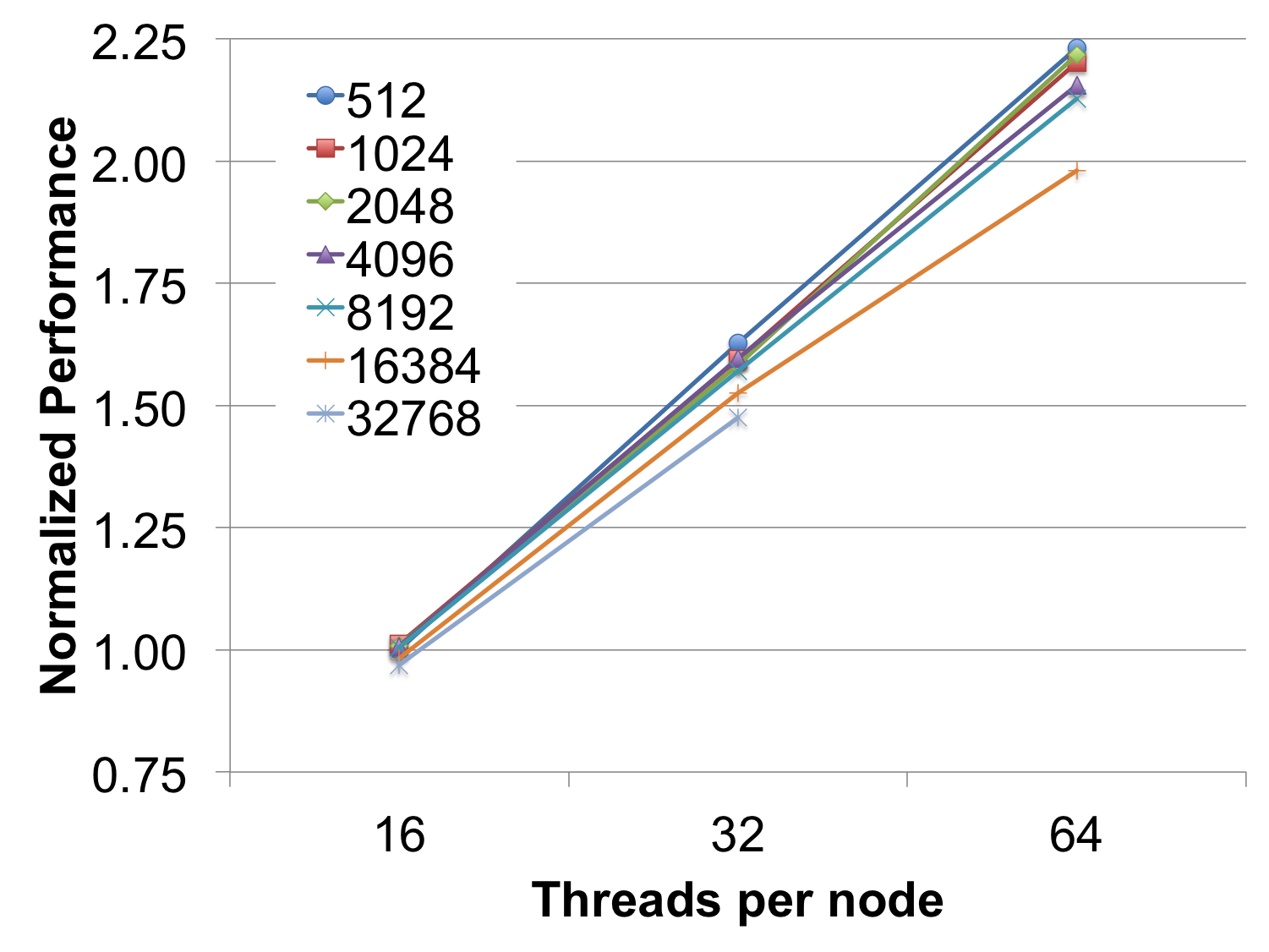}
\includegraphics[width=0.48\textwidth]{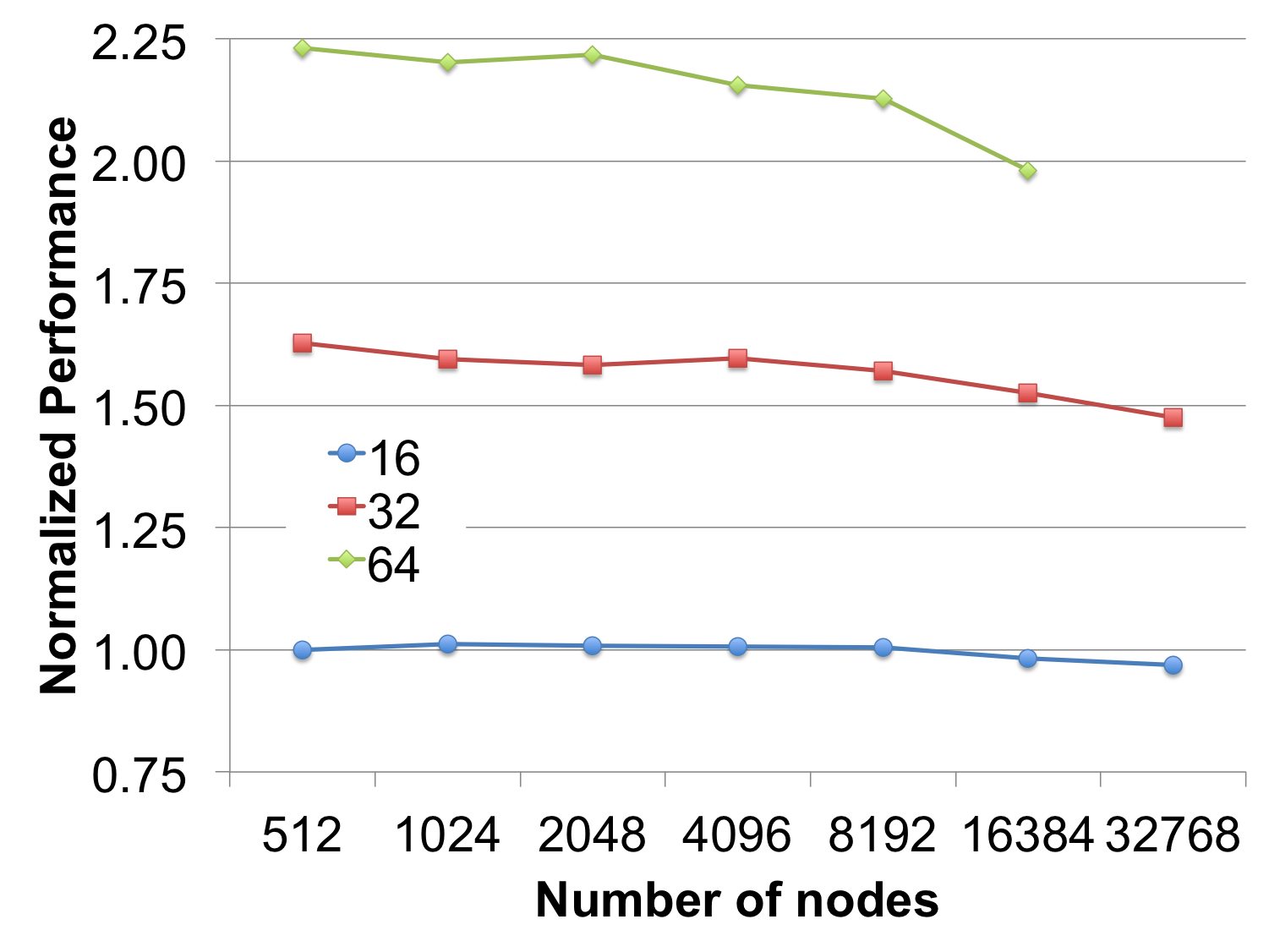}
\caption{(left) Normalized performance of \emph{Alpgen-v3} as a function of the number of ranks running on each compute node. The curve is shown for jobs with a varying total number of compute nodes. (right) Normalized performance of \emph{Alpgen-v3} jobs as a function of the total number of compute nodes. The curves correspond to different number of Alpgen ranks running on each compute node. All jobs are $W$+5jets running 450,000 weighted events per rank.}
\label{fig:perf}
\end{figure}

Figure~\ref{fig:perf} shows the performance of \emph{Alpgen-v3} as a function of threads per node and number of compute nodes. These jobs represent the performance for generating $W$+5jets with each rank generating 450,000 weighted events. The $512\times16$ configuration is used as a reference; therefore, all points are normalized to the performance of this configuration, $P/P_{512\times16}$. 

%Having made no algorithmic changes to Alpgen, the performance achieved at large parallel scales is impressive. The data shows the best performance at $512\times64$ which is 223\% more efficient than the baseline, with slowly decreasing performance as the number of nodes increases. The largest job, $16,384\times64$ or 1 million threads, gives 198\% better performance than the baseline, down 25\% from the best. This data motivates the use of only smaller jobs. Although, the run-time for generating large sets of data, for instance a 1\,fb$^{-1}$ $W$+5jets sample, is 30 minutes at $16,384\times64$ as compared to more than 13 hours at $512\times64$. The maximum job duration on Mira is 12 hours allowed for 512-node jobs on Mira. 
Having made no algorithmic changes to Alpgen, the performance achieved at large parallel scales is impressive. The data shows the best performance at $512\times64$ which is $2.2\times$ more efficient than the baseline, with slowly decreasing performance as the number of nodes increases. The largest job, $16,384\times64$ or 1 million threads, gives $2.0\times$ better performance than the baseline, down 11\% from the best. If core-hour cost is the driving factor this data motivates the use of smaller jobs. However, for LHC event generation, throughput is more important and current queuing policies on Mira give larger jobs higher priority. Therefore, the authors typically run in $16,384\times64$ configurations to achieve a higher throughput at the expense of a small decline in efficiency. The job run-time is another consideration for generating large sets of data, for instance a 1\,fb$^{-1}$ $W$+5jets sample, is 30 minutes at $16,384\times64$ as compared to more than 13 hours at $512\times64$. The maximum job duration on Mira is 12 hours for 512-node jobs on Mira. 

\begin{figure}
\centering
\includegraphics[width=0.48\textwidth]{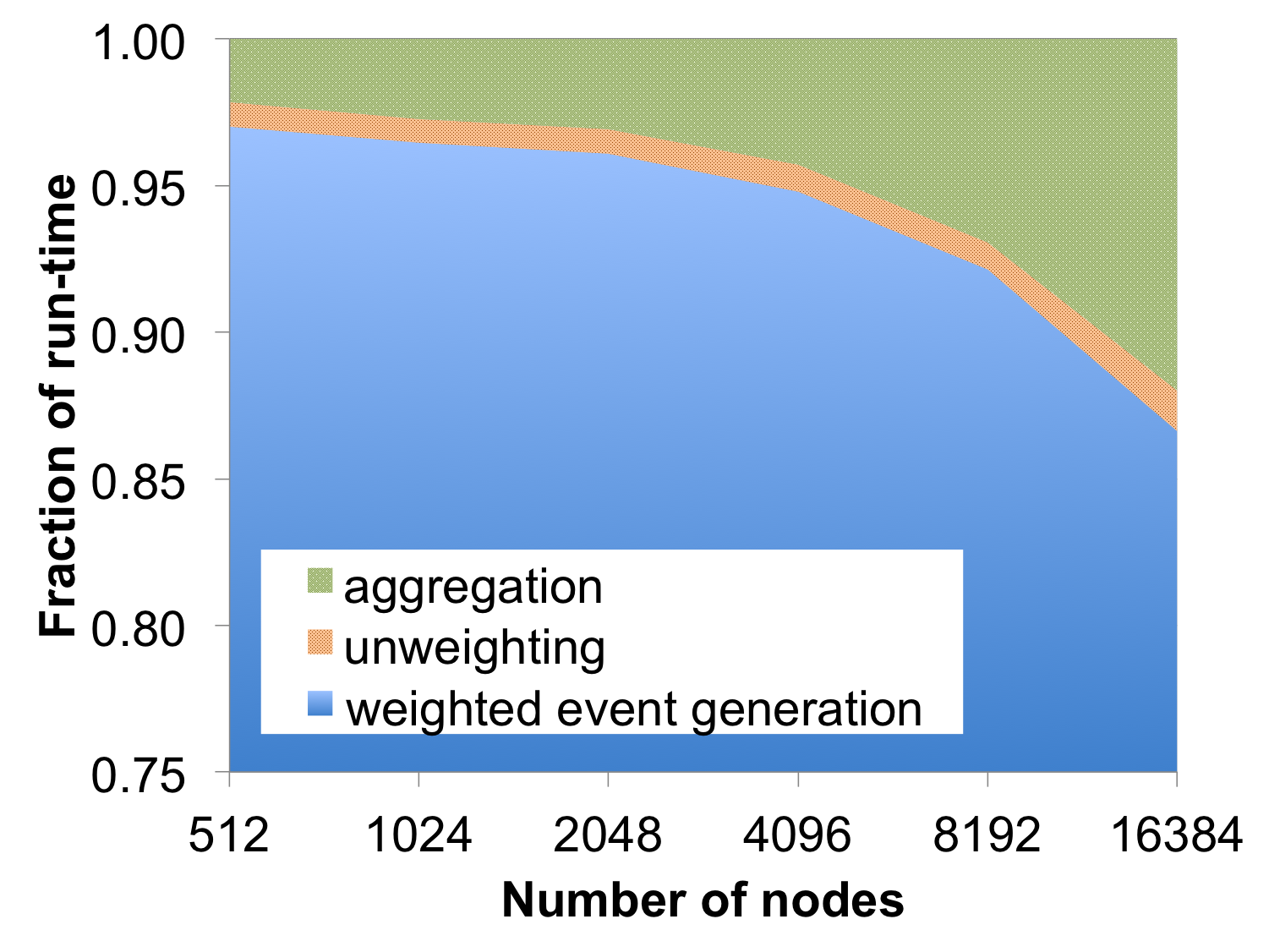}
\caption{ The fraction of the total run-time taken by the weighted event generation, unweighting, and aggregation steps at 64 threads per Mira node of \emph{Alpgen-v3} for different numbers of nodes. All jobs are $W$+5jets running 450,000 weighted events per rank.}
\label{fig:run_time_fraction}
\end{figure}

The performance begins to decline beyond 8,192 nodes. Figure~\ref{fig:run_time_fraction} shows how the fraction of run-time changes with increasing job size at 64 threads per node. The aggregation is the limiting factor because \emph{Alpgen-v3} combines all unweighted event files to a single file. To do this, MPI communicates the per-rank data to a single rank to be written to the filesystem limiting throughput. This could be optimized by aggregating to some subset of files, e.g. writing one file per some number of ranks. The increase in the aggregation run-time as the job size increases is responsible for the performance difference between the  $512\times64$ and $16384\times64$ configurations in Figure~\ref{fig:perf}.

\section{Conclusion}

Event generation is essential for LHC experimental studies. With Run II currently underway, and the high-luminosity upgrade planned for Run III, an increase in computing requirements of an order of magnitude is expected, outpacing the expected growth in capacity of the Grid. Meanwhile, the trend in leadership computing is toward orders of magnitude larger systems in the coming decade. It is, therefore, strategic to adapt codes to run on these systems to deliver increased capacity today and on future systems.
This work describes adaptation of the serial application Alpgen to run as a large-scale parallel application on Mira. By introducing MPI to improve data management, combining multiple phases of application execution in a single batch job, and utilizing compute-node RAM disks to store temporary data, Alpgen scaled to over a million threads.

\section{Acknowledgments}

This work is supported by the U.S. Department of Energy, Office of Science , Basic Energy Sciences, under contract DE-AC02-06CH11357. This research used resources of the Argonne Leadership Computing Facility, which is a DOE Office of Science User Facility supported under Contract DE-AC02-06CH11357. An award of computer time was provided by the DOE Office of Advanced Scientific Computing Research (ASCR) Leadership Computing Challenge (ALCC) program.

\section*{References}

\bibliographystyle{ieeetr}
\bibliography{main}

\end{document}